\documentclass[twocolumn]{IEEEtran} 
\usepackage{graphicx}

\def\BibTeX{{\rm B\kern-.05em{\sc i\kern-.025em b}\kern-.08em
    T\kern-.1667em\lower.7ex\hbox{E}\kern-.125emX}}

\begin{document}

\title{A High-Level Reconfigurable Computing Platform Software Frameworks} 

\author{Darran Nathan, Kelvin Lim Mun Kit, Kelly Choo Hon Min,\\Philip Wong Jit Chin, Andreas Weisensee \thanks{The authors are with the
DSP Technology Centre, School of Engineering, NgeeAnn Polytechnic, Singapore. (e-mail: darran@projectproteus.org [Darran Nathan]).}}

\markboth{Project Proteus}
{Murray and Balemi: Using the Document Class IEEEtran.cls} 

\maketitle

\begin{abstract}
Reconfigurable computing refers to the use of processors, such as Field Programmable Gate Arrays (FPGAs), that can be modified at the hardware level to take on different processing tasks. A reconfigurable computing platform describes the hardware and software base on top of which modular extensions can be created, depending on the desired application. Such reconfigurable computing platforms can take on varied designs and implementations, according to the constraints imposed and features desired by the scope of applications. This paper introduces a PC-based reconfigurable computing platform software frameworks that is flexible and extensible enough to abstract the different hardware types and functionality that different PCs may have. The requirements of the software platform, architectural issues addressed, rationale behind the decisions made, and frameworks design implemented are discussed.
\end{abstract}


\section{Introduction}
\label{sectIntroduction}
Computer processors have for many years been designed based on the von-Neumann or Harvard architectures. Software to be run on these processors are compiled into a set of processor-specific instructions, which are loaded during run-time and executed sequentially. Such sequential processing of an instruction every few clock cycles works well enough for typical PC applications such as text editors, which have low data processing requirement.

However, PCs are also often used for computationally intensive high-throughput data processing, especially in scientific research work. The sequential nature of the typical PC processor, such as the Intel Pentium, becomes a major processing bottleneck in such situations. The solution to this problem has been to use processors with greater clockspeeds, or to network several of these PCs together into a cluster or computational grid \cite{bibGlobus}.

More recently, there has been an increasing interest in the use of reconfigurable hardware chips for such computationally and data intensive processing. These chips, such as Field Programmable Gate Arrays (FPGAs), possess a fundamentally different architecture from the typical von-Neumann or Harvard type processors. The algorithms to be executed are normally defined in a hardware description language and compiled into a bitstream, which will be downloaded to the FPGA as and when use of the algorithm is desired. This bitstream download will reconfigure the hardware logic on the FPGA accordingly, allowing data passed into the FPGA to be processed in hardware, in parallel.

Several reconfigurable computing research projects \cite{bibRaw} \cite{bibPipeRench} \cite{bibGarp} focus on developing new, improved designs of reconfigurable chips. Other groups \cite{bibUMASS} \cite{bibToronto} \cite{bibBYU} utilize off-the-shelf FPGAs, such as those from Xilinx \cite{bibXilinx}, and work on issues such as logic placement and routing optimization \cite{bibVPR}. Project Proteus \cite{bibProjectProteus} was initiated by the DSP Technology Centre of NgeeAnn Polytechnic (Singapore) to develop a low-cost FPGA-based reconfigurable computing platform for typical PCs, with off-the-shelf hardware components and a portable software platform layer, that is flexible and extensible enough to abstract the different hardware types and functionality that different PCs may have. This paper discusses the requirements and design of this software platform.

Section \ref{sectRequirements} describes the requirements of the Proteus Software Platform, Section \ref{sectArchitecture} discusses the architectural issues addressed and the design of the software platform, Section \ref{sectDeployment} explains how the software platform deploys algorithms to available hardware, and finally Section \ref{sectConclusion} concludes this paper.

\section{Requirements of the Proteus Software Platform}
\label{sectRequirements}
To understand the architectural design of the software platform, it will be useful to first discuss the requirements imposed by the desired use and level of flexibility of the platform.

Firstly, the goal of the project has been to develop a PC-based reconfigurable computing platform. PCs run a variety of operating systems (OS), such as Microsoft Windows and Linux. It is therefore desirable for the software platform to be portable across various OS environments.

Secondly, being PC-based also brings the advantage of being able to utilize the various PC resources, such as plentiful RAM and harddisk storage space, and network connectivity. The software platform must be able to abstract access to sink / source data from these resources. On top of that, there must also be the possibility of using several FPGA chips concurrently (which may exist on several different PCI boards).

Thirdly, the high level of variability of available numbers and types of PC resources as well as reconfigurable processors means that the software platform has to be highly modular, with hardware abstraction modules that can be dynamically loaded according to the available resources.

Fourthly, this wide resource variation also has an implication on the deployability of algorithms - certain algorithm implementations may be suitable for execution only on certain processor types, eg) a reconfigurable hardware bitstream compiled for a Xilinx Virtex FPGA cannot be downloaded to an Altera \cite{bibAltera} Stratix FPGA, though both chips may exist in the same PC. The software platform will therefore have to match the available hardware types with the available compatible algorithm implementations.

Finally, all this need for flexibility in the software platform of being able to load different hardware abstraction and algorithm implementation modules means that such modules should be easily created in a high-level language that most programmers are familiar and comfortable with.

\section{Architecture of the software platform}
\label{sectArchitecture}
Considering the requirements set out in Section \ref{sectRequirements}, a high-level and modular software platform frameworks was designed.

The requirements for portability across OS environments, modularity of extensions, and ease of programmability, led to the Java language being selected for implementation of the software platform. This allows the software platform to be run on any computer that has a Java Virtual Machine (JVM) installed, while the high-level and object-oriented nature of the language satisfies the requirements of dynamically loadable modules that can be easily programmed in a widely-adopted language.

\begin{figure}[htb]
\begin{center}
\includegraphics[width=0.2\textwidth]{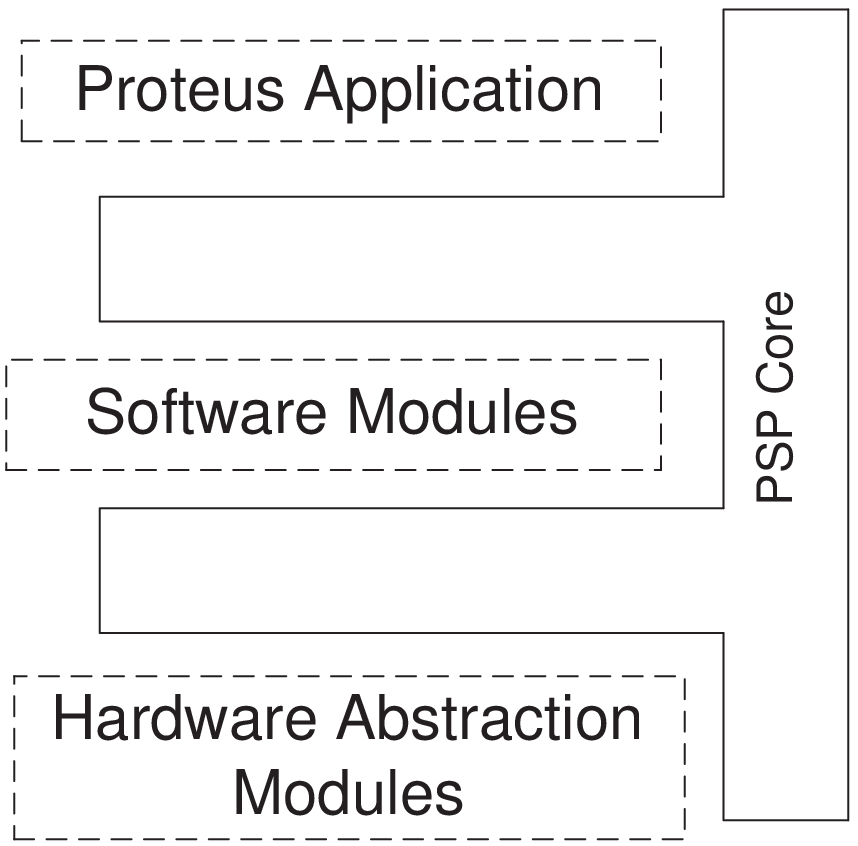}
\caption{Components of the Proteus Software Platform}
\label{figComponents}
\end{center}
\end{figure}

To modularize its functionality, the software platform has been divided into four main component blocks: the Proteus Software Platform (PSP) core, which holds the common set of interfaces and functionality, and three other components: the Proteus Application, Hardware Abstraction Modules (HAMs), and Software Modules, that are deployed according to the available functionality on the PC and the desired application. The use of Java allows each of these component modules to be distributed as individual JAR files. This segmentation is illustrated in Figure \ref{figComponents} and described in greater detail below.

\subsection{Software Modules}
\begin{figure}[htb]
\begin{center}
\includegraphics[width=0.5\textwidth]{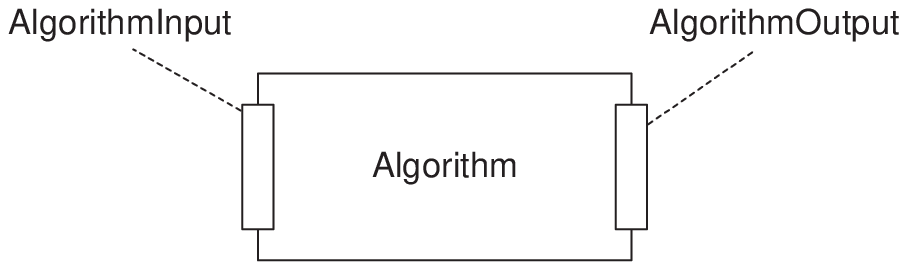}
\caption{A typical algorithm block representation}
\label{figAlgoBlock}
\end{center}
\end{figure}

An Algorithm block defines a unit of operations that receives data at an input, processes it, and sends the results out through an output. This is commonly represented by a block as shown in Figure \ref{figAlgoBlock}.

Each of these blocks is usually of a processor-specific implementation, such as a compiled Java class, or an FPGA hardware implementation bitstream.

However, the Proteus Software Platform is intended to be run in environments where the available processor types are variable and determined only during run-time, and where Algorithms may have a number of implementations for different processor types.

Hence there is a need for a different Algorithm structure, one which allows for a high level description of the connectivity between Algorithm blocks, while allowing each block to have multiple implementations for the various processor types.

\begin{figure}[htb]
\begin{center}
\includegraphics[width=0.5\textwidth]{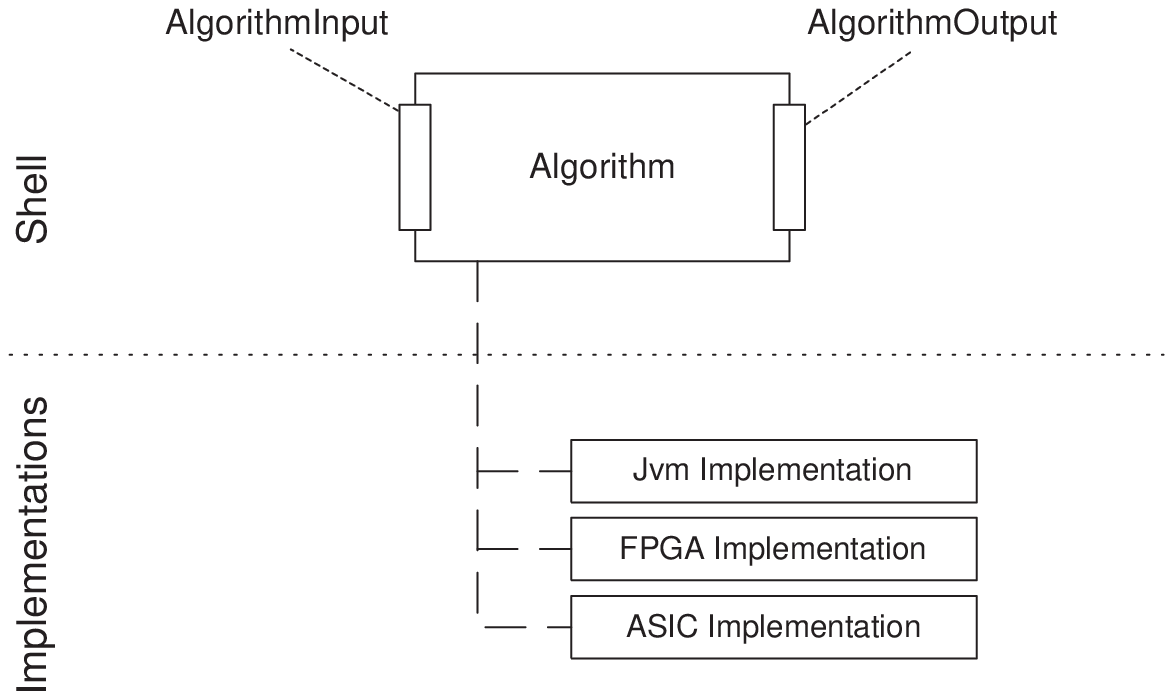}
\caption{The Algorithm 'shell/implementation' structure}
\label{figAlgoShellImp}
\end{center}
\end{figure}

The resulting design takes on a 'shell/implementation' architecture, as shown in Figure \ref{figAlgoShellImp}. In this structure, the Algorithm 'shells' are connected up to one another, and define the input/output data types. A 'shell' can be associated with multiple 'implementations', each of which is compatible with a different processor type (such as an FPGA or JVM).

\begin{figure}[htb]
\begin{center}
\includegraphics[width=0.5\textwidth]{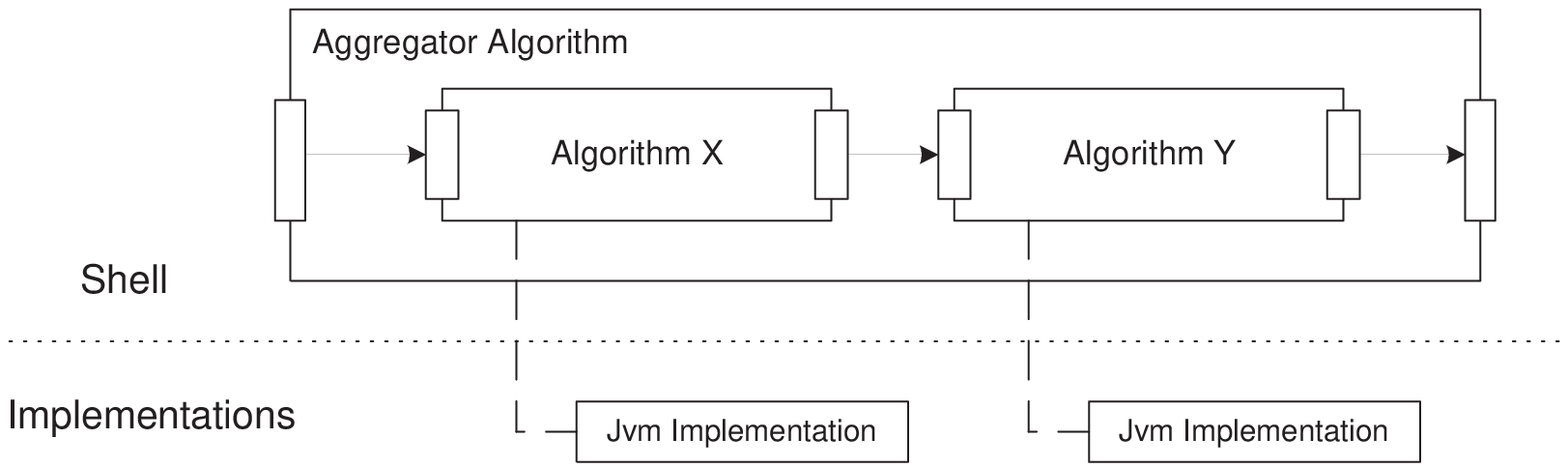}
\caption{Connecting up a number of Algorithm 'shells' to form a data flow graph}
\label{figAlgoAggregate}
\end{center}
\end{figure}

Connecting up a number of 'shells' will therefore create a high level data flow graph, ensuring that data will be passed correctly from one algorithm to the next, independent of where the associated 'implementations' are deployed. This is illustrated in Figure \ref{figAlgoAggregate}.

\subsection{Hardware Abstraction Modules (HAMs)}
The need for the ability of the software platform to utilize various kinds of processor types and other PC resources implies a need to define a common layer of abstraction to all these resources. This abstraction layer must provide information on the type of Algorithm implementations that are compatible with corresponding physical hardware, as well as whether a compatible Algorithm implementation can be deployed to that hardware (eg, if the processor is not already overloaded).

\begin{figure}[htb]
\begin{center}
\includegraphics[width=0.5\textwidth]{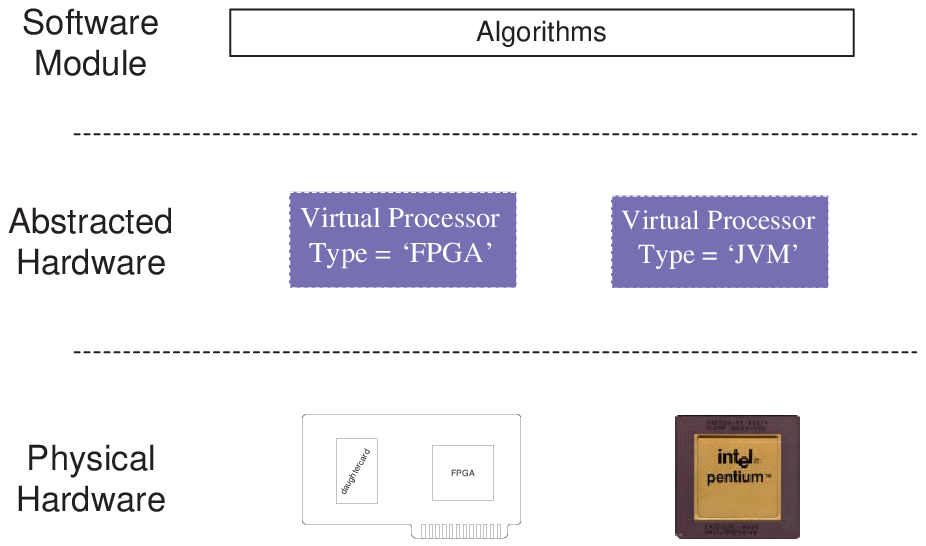}
\caption{Abstraction of physical hardware via 'Virtual Processors'}
\label{figHam}
\end{center}
\end{figure}

The abstraction layer designed to satisfy the above requirements consists of modelling the desired properties of one or more physical hardware resources in one or more 'virtual processor' entities, as shown in Fig \ref{figHam}.

These 'virtual processors' will be queried by the software platform to determine the compatibility and deployability of a particular Algorithm implementation, as described in detail in Section \ref{sectDeployment}.

For a particular physical hardware resource (such as an FPGA processor board, or a storage media), the 'virtual processor' is part of a larger package called the 'Hardware Abstraction Module' (HAM), which is a distribution JAR of all the hardware-specific components (such as the interfaces to the software platform, and the OS-specific device drivers).\subsection{Proteus Application}
The Proteus Application serves two purposes - it provides an administrative interface to the end-user, and defines the mechanism by which data is passed from one Algorithm to another.

The administrative interface allows the end-user to perform such operations as starting / stopping the platform or selecting the desired algorithm for download.

The data passing mechanism is defined in the Proteus Application because various techniques exist, such as Communicating Sequential Processes (CSP) \cite{bibCSP} and Dataflow Process Networks (PN) \cite{bibPN}, and utilization of a particular mechanism is application-dependent.

\begin{figure}[htb]
\begin{center}
\includegraphics[width=0.5\textwidth]{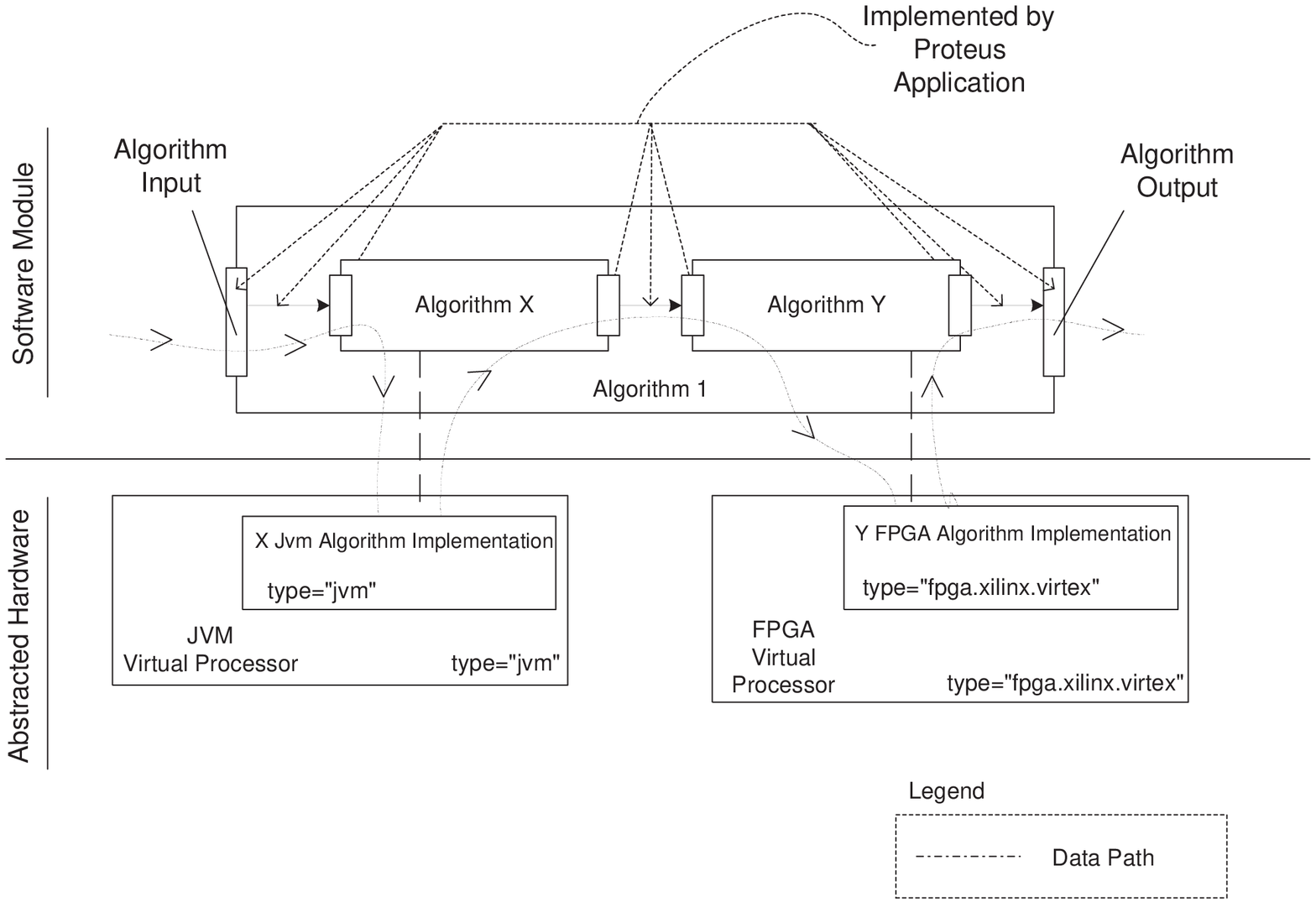}
\caption{Data exchange mechanism implemented by Proteus Application}
\label{figDataExchange}
\end{center}
\end{figure}

Figure \ref{figDataExchange} shows the set-up of Processors and Algorithms, with the AlgorithmImplementations deployed to corresponding Processors. The portions concerning data-exchange, which have to be implemented by the Proteus Application, are marked accordingly.

\section{Deployment of Algorithms}
\label{sectDeployment}
For the software platform to perform the tasks of matching Algorithm implementations with virtual processors, a technique of tagging both of these with some common form of type compatibility identification is needed. This tagging should offer the ability to define different levels of compatibility, such as that at a specific chip model or at a higher family level. For example, an FPGA Algorithm implementation may be compatible with only the Xilinx Virtex XCV100 chip, or may be compatible with all chips in the Virtex family, and should be allowed to be tagged as such.

\begin{figure}[htb]
\begin{center}
\includegraphics[width=0.5\textwidth]{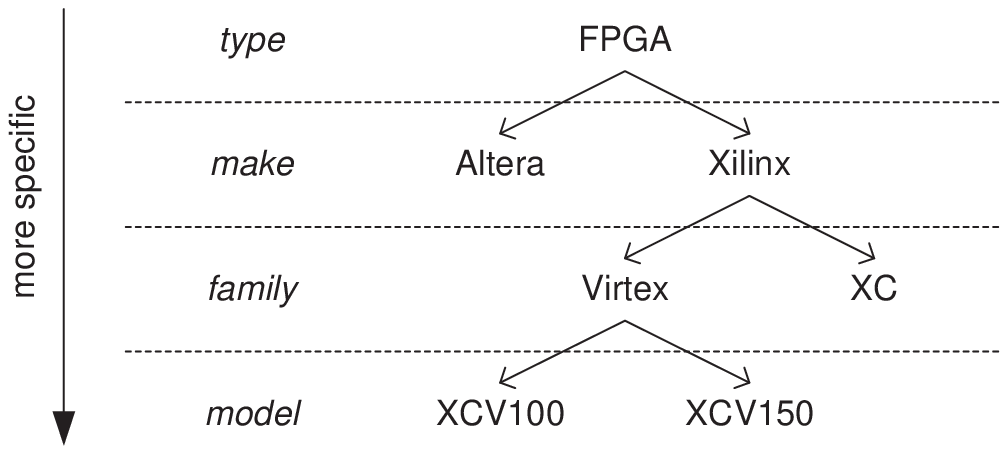}
\caption{Algorithm implementation / virtual processor tagging scheme}
\label{figTree}
\end{center}
\end{figure}

The tagging mechanism designed consists of a string in the general form "type.make.family.model.otherInfo", that can have any number of descriptors separated by dots ("."), depending on the level at which an Algorithm implementation or virtual processor is specific. For example, an Algorithm implementation that can be downloaded to a Xilinx Virtex series XCV100 chip may be tagged "fpga.xilinx.virtex.xcv100", while a virtual processor that accepts all Xilinx Virtex Algorithm implementations may have that of "fpga.xilinx.virtex". The specificity of a tag increases with the number of descriptors. Such a scheme can be illustrated as a tree graph, as shown in Figure \ref{figTree}.

In this tree graph, the least specific descriptor is at the top - in this case the 'type' level, with the descriptor value "FPGA". Moving down one level introduces the next more specific 'make' descriptors, so a tag here may be "FPGA.Xilinx". When the Proteus Software Platform tests whether an Algorithm implementation is compatible with a virtual processor, it needs to only ensure that the tag of the virtual processor is located at the same point on the tree, or is an ancestor of that of the Algorithm implementation. That is, a more specific (lower in the tree) Algorithm implementation can only be deployed to an equal or less specific virtual processor (equal or higher in the tree). For example, an Algorithm implementation tagged "FPGA.Xilinx.Virtex.XCV100" is compatible with a virtual processor of type "FPGA.Xilinx.Virtex", but not "FPGA.Xilinx.Virtex.revB".

\begin{figure}[htb]
\begin{center}
\includegraphics[width=0.5\textwidth]{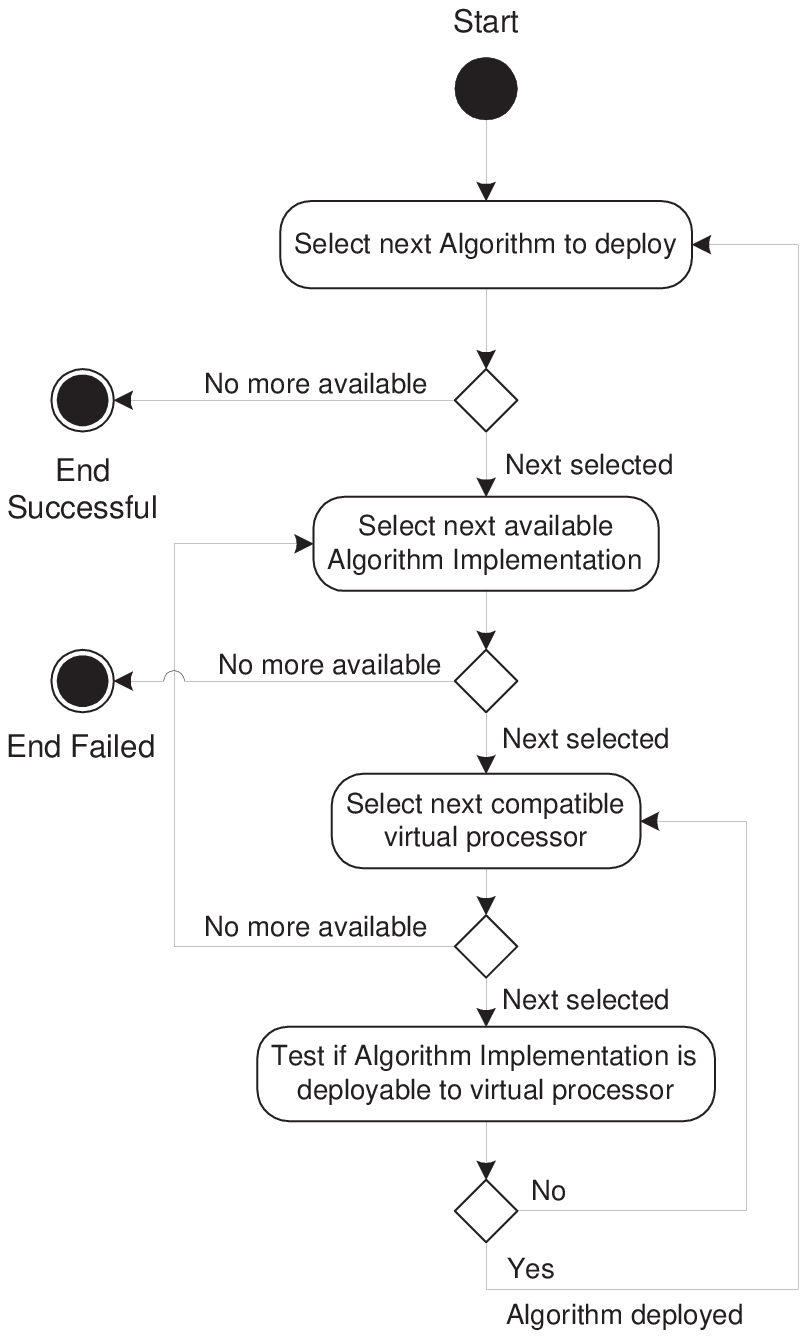}
\caption{Algorithm deployment flow}
\label{figFlow}
\end{center}
\end{figure}

For each Algorithm to be deployed, the software platform runs through the list of available Algorithm implementations and virtual processors to identify those that are compatible. Once a match is found, the virtual processor is queried if the matching Algorithm implementation can be deployed to it. This deployability step is necessary to test if the associated hardware has the necessary capacity to run the compatible Algorithm implementation, e.g. if an FPGA has sufficient available space. If not, the process is repeated till a match that is both compatible and deployable is found. This flow for Algorithm deployment is illustrated in Figure \ref{figFlow}.

\section{Conclusion}
\label{sectConclusion}
The software platform designed satisfies the requirements for a high-level, portable reconfigurable computing platform frameworks that is highly modular and flexible enough to utilize the varied resources available on different PCs.

\section*{Acknowledgments}
We gratefully acknowledge the funding support provided by the NgeeAnn Kongsi (Singapore) and NgeeAnn Polytechnic's Innovation \& Enterprise Office.

\nocite{*}
\bibliographystyle{IEEE}

\begin{thebibliography}{1}

\bibitem{bibGlobus}
Globus Alliance
\newblock {\em http://www.globus.org}

\bibitem{bibRaw}
RAW Project, Laboratory for Computer Science, MIT
\newblock {\em http://cag-www.lcs.mit.edu/raw/}

\bibitem{bibPipeRench}
PipeRench Project, Carnegie Mellon University
\newblock {\em http://www.ece.cmu.edu/research/piperench/}

\bibitem{bibGarp}
Garp Project, BRASS Research Group, UC Berkeley
\newblock {\em http://brass.cs.berkeley.edu/garp.html}

\bibitem{bibUMASS}
Reconfigurable Computing Group, University of Massachusetts
\newblock {\em http://www.ecs.umass.edu/ece/tessier/rcg/}

\bibitem{bibToronto}
FPGA Research Group, University of Toronto
\newblock {\em http://www.eecg.toronto.edu/EECG/RESEARCH/FPGA.html}

\bibitem{bibBYU}
BYU Configurable Computing Laboratory
\newblock {\em http://splish.ee.byu.edu/projects/projects.html}

\bibitem{bibXilinx}
Xilinx Inc.
\newblock {\em http://www.xilinx.com}

\bibitem{bibVPR}
VPR: Versatile Packing, Placement and Routing for FPGAs
\newblock {\em http://www.eecg.toronto.edu/~vaughn/vpr/vpr.html}

\bibitem{bibProjectProteus}
Project Proteus, DSP Technology Centre, NgeeAnn Polytechnic, Singapore
\newblock {\em http://www.projectproteus.org}

\bibitem{bibAltera}
Altera Inc.
\newblock {\em http://www.altera.com}

\bibitem{bibCSP}
C.A.R. Hoare
\newblock {\em Communicating Sequential Processes}
\newblock Prentice Hall, 1986

\bibitem{bibPN}
Edward A. Lee, Thomas M. Parks
\newblock {\em Dataflow Process Networks}
\newblock Proceedings of the IEEE, vol. 83, no. 5, pp. 773-801, May 1995 

\end{thebibliography}

%


%

\end{document}